\documentstyle[12pt,special,dina4]{article}
\begin{document}
\setlength{\parindent}{0cm}
\normalbaselineskip 17pt
\normalbaselines
\def\dps{\displaystyle}
\def\intuu{\int\limits_\infty^\infty}
\def\nex{\\[10pt]}
\def\uex{U^{ex}}
\def\intl{\int\limits}
\def\en{\end{equation}}
\def\vec#1{\underline{#1}}
\def\op#1{\hat{\rm #1}}
\def\vac{\mid 0\!\!>}            % def vacuum
\def\fsd{f_\sigma^\dagger}   % def erzeuger
\def\fmsd{  f_{m\sigma}^\dagger}
\def\fms{  f_{m\sigma}}
\def\cskd{c_{\sigma\underline{k}}^\dagger}
\def\fs{ f_\sigma}           % def vernichter
\def\csk{c_{\sigma\underline{k}}}
\def\sums{\sum_\sigma}       %
\def\sumsk{\sum_{\sigma\underline{k}}}
\def\limb{\lim_{\beta \rightarrow \infty}}
\def\limbb{\limb \frac{1}{\beta}}
\def\iwn{i \omega_n}
\def\iw#1{i \omega_{n_{#1}}}
\def\intinf{\intl_{-\infty}^{\infty}}
\def\e{\epsilon}
\def\g{\mid\! g\!>}
\def\w{\omega}
\def\oin{\oint_{\cal C}}
\def\oiz{\oin\frac{dz}{2\pi i}}
\def\oiw{\oin\frac{d\w}{2\pi i}\frac{e^{-i\w t}}{\w^2}}
\def\oizz{\oin\frac{dz}{2\pi i}e^{-\beta z}}
\def\oib{\oiz e^{-\beta z}}
\def\bge{\begin{eqnarray}}
\def\ene{\end{eqnarray}}
\baselineskip 17pt
\parindent 0mm
\parskip 4mm plus 1mm minus 1 mm
\topsep 0mm
\partopsep 0mm
%   text-gleitobjekte  anst"ande
\textfloatsep 10mm plus 2mm minus 4mm
\intextsep 6mm plus 2mm minus 2mm
\floatsep 10 mm plus 2mm minus 2mm
\def\ul#1{\underline{#1}}
\def\ek{\epsilon_{\ul{k}}}
\def\ekk#1{\epsilon_{\ul{k_{#1}}}}
\def\kk#1{\ul{k_{#1}}}
\newcommand{\ho}[1]{$^{#1}$}
\newcommand{\ti}[1]{$_{#1}$}
\pagenumbering{arabic}
\title{Perturbational approach to the Anderson Model: \\
New results from a
post-NCA treatment}
\author{F.B.~Anders and N.~Grewe \\
Institut f\"ur Festk\"orperphysik, \\
Technische Hochschule D-64289 Darmstadt, 
Germany}
\date{April 1994}
\maketitle

\begin{abstract}
We present selected results from combined analytical and numerical studies of
the Anderson-impurity model within the framework of infinite order perturbation
theory with respect to the hybridization. Our approximation goes considerably
beyond the well known non\-cross\-ing approximation (NCA): The re-summations include
skeleton diagrams up to infinite order of the cross\-ing variety, 
including all
$1/N^2$ contributions ($N =$ degeneracy of the magnetic state). We demonstrate,
in a comparison with NCA theory, changes in the threshold behaviour of local
propagators, a qualitative improvement of spectral properties and a clear 
progress in attaining local Fermi-liquid properties. Further applications of
the post-NCA theory are pointed out.
\end{abstract}


\vspace{5mm}
Accepted for publication in Europhysics Letters

\newpage
More than ten years ago the first evaluations of a self-consistent infinite
order perturbational theory of the Anderson impurity model with respect
to the hybridization had been reported \cite{GREKURA83}. 
The approach was named non-crossing approximation (NCA), because all
contributions included in the self-energy of local propagators result from just
one skeleton diagram (per local state) which generates all processes without 
crossing band-electron lines. Viewed from an expansion scheme in the inverse of
a degeneracy $N$ of the magnetic local state the NCA is the leading 
self-consistent approximation of order $1/N$. It furnishes in particular the
correct low-energy scale of the problem \cite{MUELLERH84}, Wilson's 
Kondo temperature $T_K$. 
Due to an imbalance of particle and hole excitation processes 
\cite{KEITER85} the
NCA contains a pathology at very low energies $T_{path} \ll T_K$, which e.g. 
shows up as a residual spike in one-particle spectra at the Fermi energy
$\mu = \epsilon_F = 0$. 
Moreover, the spectral weight accumulated in form of the
``Abrikosov-Suhl (AS) resonance'' near the Fermi level via the Kondo mechanism
is subject to deformations within NCA: These deformations are particularly
strong for low degeneracy, e.g. $N = 2$, and in the intermediate valence regime
where charge-fluctuation processes become important. They result in a violation
of exact fundamental properties, such as the Friedel sum rule or even 
positivity-requirements for self-energies. Since these deficiencies can be
isolated and controlled \cite{MUELLERH84} 
and moreover become negligible for whole ranges 
of excitation energies or in the large-$N$ limit, the NCA has developed as a 
valuable tool in the study of strongly correlated electrons and has served for 
a number of applications even in the lattice problems \cite{GREW87}. 
It also compares favourably to other many-body techniques, 
even exact (numerical) ones \cite{BONCA};
lately it has been used successfully for the Hubbard model \cite{PRU90}
and also in the limit of large dimension \cite{PRUCOXJAR93}.

In spite of its simple and intuitive form the NCA is not easily solved.
Analytical solutions for a few quantities like threshold exponents or 
low-energy propagators exist for $T = 0$ \cite{MUELLERH84}, 
but otherwise a numerical 
solution of two coupled singular integral equations is necessary. Originally,
even the limit of infinite local Coulomb repulsion $U$ had to be assumed. An
extension to the range $U < \infty$ with good results has proven possible at
considerable calculational expense \cite{PRUQIN89}. 
Up to recently it did not seem feasible
to include skeleton diagrams of sixth and higher order (necessarily with 
crossing band-electron lines) in the numerical solution, which would be 
necessary e.g. to go consistently to the next order in the large-$N$ expansion
scheme. Although the corresponding analytical formulation of the 
self-consistency equations, even with inclusion of further 
Coulomb-matrix elements like direct exchange, has been published 
recently \cite{AGQ91}, the multiple integration problem
caused by the more complicated electron-hole excitations in the self-energy
diagrams proved too difficult. By a reformulation of the self-consistency
problem of this post-NCA theory (PNCA) in terms of a vertex correction
function, also to be determined self-consistently, one of us \cite{ANDERS94}
has opened
the way to a successful evaluation at infinite $U$. Since the results so far
convincingly demonstrate the qualitative improvement reached by the new theory
we think it justified to present in the following a few selected key quantities
which are frequently used in discussions of systems with strong electron
correlations \cite{GRESTEG90}.
 
For brevity we summarize the analytical content of our PNCA in the well known
diagrammatic language used in this field \cite{KK71,BICKERS} in fig.~1
reducing the equations of ref.~\cite{AGQ91} 
to the case of vanishing direct exchange I. Results for $I > 0$ and a
more extensive general discussion will be published elsewhere. The equations
coupling local self-energies and vertex correction self-consistently up to 
order $1/N^2$ (and non systematically higher) are:
\begin{equation}
\label{equ-1}
\begin{array}{lcr}
\Sigma_0(z) &=&{\dps\sum_{m=1,N}} |V|^2\intuu de\rho(e)f(e) P_m(z+e)\Delta^{I}_m(z,z+e)\nex
\Sigma_m(z) &=& \phantom{\sum_{M=1,N}}|V|^2\intuu de\rho(e)f(-e) P_0(z-e)
\Delta^{O}_m(z,z-e)
\end{array}
\end{equation}
$\rho$ denotes the density-of-states of unperturbed conduction electrons.
In the absence of an external magnetic field the ionic propagators $P_m$ 
are identical. Hence, also using the obvious symmetry concerning the 
energy arguments the vertex corrections reduce to
\begin{equation}
\label{equ-delta-pm}
\begin{array}{l}
\Delta^{I}_m(x,y) = \Delta^{O}_m(y,x) \equiv 
\dps \Delta(x,y) \; = \nex
\begin{array}{cr}
1\; - & \dps |V|^4\intuu\intuu dudv \rho(u)\rho(v)f(u)f(-v)
 \mbox{G}(x,x+u,x+u-v)\nex
&\cdot \Delta(x+u-v,y+u-v) \mbox{H}(y,y-v,y+u-v)\nex
\nex
\end{array}
\end{array}
\end{equation}
using the auxiliary functions
\begin{equation}
\label{equ-aux-vari-1}
\begin{array}{l}
%
% begin G
%
\mbox{G}(x,x+u,x+u-v)\; = \\
\Delta(x,x+u) P_m(x+u) \Delta(x+u-v,x+u)P_0(x+u-v) 
\nex
\end{array}
\end{equation}
\begin{equation}
\label{equ-aux-vari-2}
\begin{array}{l}
%
% begin 
%
\mbox{H}(y,y-v,y+u-v)\; = \\
P_m(y+u-v)\cdot \\
\left(\Delta(y-v,y+u-v)P_0(y-v)   \Delta(y-v,y)
-\mbox{K}(y+u-v,y,y+u)
\right)\nex
%
% begin K
%
\mbox{K}(y+u-v,y,y+u)\; = \\
 \begin{array}{r}\dps
N |V|^2\intuu dl \rho(l)f(-l) 
\Delta(y+u-v-l,y+u-v)P_0(y+u-v-l)\\
\cdot \Delta(y+u-v-l,y+u-l) P_{m}(y+u-l) \\
\cdot \Delta(y-l,y+u-l) P_0(y-l) \Delta(y-l,y) \nex
\end{array}
\end{array}
\end{equation}
Vertex corrections have to be included also in the local one-particle
propagator 
\begin{equation}
\label{equ-2}
G_{fm}(\iwn) = \frac{1}{Z_f}\oizz P_0(z)P_m(z+\iwn)\Delta_m(z,z+\iwn)
\end{equation}
and, however, with a different analytical structure
in an analogous fashion in the local susceptibility $\chi(\omega)$ 
\cite{KK71,BICKERS,AGQ91,ANDERS94}. They give very important 
contributions, not only implicitly in the
determination of the local propagators, but also through their explicit
appearance in eq.~(\ref{equ-2}); the latter effect is reduced in the case of
$\chi (\omega)$ since the sixth order skeleton does not contribute in this
case. We brought a numerical iteration procedure of the above system of 
integral equations to full convergence on a cluster of work stations, 
using dynamically defined logarithmic
meshes for the threefold integration in (\ref{equ-delta-pm}) 
and a number of exact sum rules \cite{KOJIMA,BICKERS,ANDERS90} 
to check its quality. 
Whilst the errors in the modulus of the propagators and the
self-energies stay within the NCA-accuracy of 2\% a major effect concerns
the validity from the Friedel sum rule: a cut-down
from 20\% error in NCA \cite{ANDERS90} to 3-6\% is found in the numerical data.
Results are presented in the following figures and
compared to corresponding NCA calculations.

For a study of the Kondo regime we choose a band of width $W = 20 \: \Delta \:
(\Delta = \pi \: V^2 \: {\cal N}_F$ being the Anderson width for hybridization
$V$ and DOS ${\cal N}_F$ at $\epsilon_F$) with constant DOS and smooth 
edge-cutoffs, an unperturbed local level position $\Delta E = - 3 \Delta$ and
fixed low degeneracy $N = 2$ in order to present the case for which the NCA 
shows maximal deficiencies. Fig.~2 presents the resolvent spectra of the empty 
(a) and occupied (b) local states for three different temperatures $T = 0.5 \:
T_K, \: 1.0 \: T_K$ and $2.0 \: T_K$. The most prominent effect lies 
in a relative
shift of the peak positions of both propagators near the common threshold, 
which are close to each other already for temperatures $T$ of order 
$T_{K}$, in contrast to the NCA. Moreover, a detailed analysis of the
corresponding slopes reveals distinct changes in the threshold exponents. These
will be the subject of a forthcoming publication.

One-particle spectra of electrons in the local state are shown in fig.~3 for 
the same three temperatures as before. 
As already expected by M\"uller-Hartmann ten years ago \cite{MUELLERH84}, 
a rather large impact of the vertex correction
$\Delta(z,z+\iwn)$ in eq.~(\ref{equ-2}) is recognised 
shifting the AS-resonance significantly towards the chemical potential 
$\mu = 0$ compared to the NCA as shown in fig.~2. This is in accord with the 
density-of-states rule, derived from the Friedel sum rule 
\cite{LANG,ANDERS90} in the limit $T\rightarrow 0$, which demands
\begin{equation}
\label{equ-3}
\limb\rho_{loc} (\omega = 0) = \frac{1}{\pi \Delta} \: \sin^2 \: \left( n_f 
\frac{\pi}{N} \right) \approx 0.96 \:\frac{1}{\pi\Delta}
\end{equation}
for the present set of parameters ($n_f \approx 0.9$). 
The peak position in the NCA is much more
temperature dependent and comes close to $\mu$ only at very low
temperatures $T \ll T_K$, where it is not reliable. This last point in
particular is apparent from the temperature-dependent height of the
AS-resonance. It clearly exceeds the unitary limit value of $1/\pi \: \Delta$ 
- compare eq.~(\ref{equ-3}) - already at $T \lsim T_K$, and
saturates at about 0.45 $\Delta$ for $T \ll 
T_K$ \cite{NOZIERGRUZAWA}.
The PNCA on the other hand violates this limit only slightly (and
not much worse at lower temperatures). As is to be expected from the nature of
the perturbational approach, no approximation can perfectly redistribute the
original sharp spike near $\mu$ generated by the class of most divergent
diagrams \cite{KEIMOR84} so that also the PNCA has a remaining deficiency. 
The essential 
progress, however, not only lies in the apparent order of magnitude 
improvement, compared to the NCA, but also in a stabilisation of the low-energy
regime as evidenced by the much smoother temperature dependence and also by the
changed threshold exponents \cite{MUELLERH84,MENGE,GRUNEN,ANDERS94}.

This point of view is further supported by results for the self-energy of local
electrons as shown in fig.~4. The interesting regime again lies within a few
energies $T_K$ around $\mu$, where the AS-resonance is formed. The 
region around the original local resonance $\Delta E$ is only slightly modified
as compared to the NCA, showing a much broadened peak of dominant spectral
weight. The formation of a local Fermi liquid
 \cite{SHIBA75,YAMADA74,YAMADA75} should - in the rigorous
theory - be connected with a minimum of the self-energy's imaginary part at
$\omega = \mu \: (= 0)$ and $T \rightarrow 0$, more precisely \cite{HEWSON}:
\begin{equation}
\label{equ-4}
\limb \Im m\: \Sigma_{M}(\omega - i \delta) 
\approx \Delta\left(1 + \frac{1}{2} \: \left(\frac{\omega}{T_K}\right)^2\right)
\: \: \: ,
\end{equation}
for $\omega \ll T_K$ which is indicated in the inset of fig.~4.
The sequence of PNCA-curves clearly demonstrates this transition to local
Fermi-liquid behaviour to a much better degree than the NCA results, also
shown for comparison. The detailed form of this change corresponds well to the
effects described above, including the remaining small violation of the
unitary lower limit $\Delta$. The overall improvement reached with the PNCA is
particularly important for a study of correlated electrons on a lattice, where
nonlocal contributions to the self-energy exactly cancel the constant part
$\Delta$ \cite{GREW87}. 
Approximations on either the local or the nonlocal part of
excitations processes therefore usually have severe consequences and must be 
corrected for when calculating lattice properties, e.g. transport quantities
\cite{COGRE88}. 
There is good reason on the basis of the present results to expect a
much better treatment of such lattice effects when the PNCA is incorporated 
into the lattice version of perturbation theory 
as for example the LNCA \cite{GREW87}.
Preliminary studies support this claim and will be reported soon.

In conclusion, the PNCA improves considerably on the known analytical 
perturbation approaches to the problem of strong local electron correlations 
as evidenced by the selected properties of spectral densities shown above. 
These examples could be extended by e.g. the magnetic susceptibility or the 
specific heat which likewise show interesting modifications towards the 
expected exact 
behaviour. A more extensive discussion and broader explication of
results will be published elsewhere. A particularly interesting perspective
develops for the study of coherent behaviour in lattice models. We plan to 
report on the Anderson lattice and the Hubbard model (at infinite $U$) soon.

{\hfill *** \hfill }

Parts of the numerical calculations have been performed on IBM RISC work 
stations at the Darmstadt Institut f\"ur Graphische Datenverarbeitung;
many thanks to A.~Mavromaras.

\newpage
{\bf \underline{Figure captions}}

\begin{itemize}
\item[\bf Fig.~1:]
Diagrammatic representation of a) the self-energy eq.~(\ref{equ-1})
and b) 
of the vertex function for an incoming band-electron eq.~(\ref{equ-2}).
In the formulas we used the symmetry $\Delta(z,z+\e) :=
\Delta^{I}_m(z,z+\e) = \Delta^{O}_m(z+\e,z)$ between the vertices for 
incoming and outgoing band-electrons. Energies are given in units of $\Delta$.

\item[\bf Fig.~2:]
Comparison of the ionic propagator in NCA and $1/N^2$-approximation (PNCA)
at temperatures $T= 0.5, 1.0, 2.0 T_K$ in the vicinity of the threshold
- which is set to zero here - 
for (a) the empty state, (b) the singly occupied state.
\item[\bf Fig.~3:]
Comparison of the local one-particle propagator spectrum in NCA and PNCA
at  temperatures $T= 0.5, 1.0, 2.0 T_K$ in the vicinity of the chemical
potential $\mu = 0$.
\item[\bf Fig.~4:]
Comparison of the imaginary part of the local one-particle propagator 
self-energy in NCA and PNCA
at the temperatures $T= 0.5, 1.0, 2.0 T_K$ in the vicinity of the chemical
potential $\mu = 0$. The inset demonstrates the 
validity of eq.~(\ref{equ-4}) in the zero temperature limit with a slightly
modified value of $\Delta < 1$ (in our dimension-less energy units).
\end{itemize}

\end{document}